\documentclass[a4paper,fleqn,usenatbib]{mnras}

\usepackage{newtxtext,newtxmath}
\usepackage{url}

\usepackage[T1]{fontenc}
\usepackage{ae,aecompl}

\usepackage{graphicx}	
\usepackage{amsmath}	
\usepackage{amssymb}	

\title[Evolution of radio luminosity functions]{Does the evolution of the radio luminosity function of star-forming galaxies match that of the star-formation rate function? }

\author[M. Bonato et al.]{Matteo Bonato$^{1,2}$\thanks{E-mail: matteo.bonato@oapd.inaf.it},
Mattia Negrello$^{3}$,
Claudia Mancuso$^{1}$,
Gianfranco De Zotti$^{1,2}$,
\newauthor
Paolo Ciliegi$^{4}$,
Zhen-Yi Cai$^{5}$,
Andrea Lapi$^{1}$,
Marcella Massardi$^{6}$,
Anna Bonaldi$^{7}$,
\newauthor
Anna Sajina$^{8}$,
Vernesa Smol\u{c}i\'{c}$^{9}$
and Eva Schinnerer$^{10}$
\\
$^{1}$SISSA, Via Bonomea 265, 34136, Trieste, Italy\\
$^{2}$INAF-Osservatorio Astronomico di Padova, Vicolo dell'Osservatorio 5, I-35122 Padova, Italy\\
$^{3}$School of Physics and Astronomy, Cardiff University, The Parade,
Cardiff CF24 3AA, UK \\
$^{4}$INAF-Osservatorio Astronomico di Bologna, Via Ranzani 1, I-40127 Bologna, Italy\\
$^{5}$CAS Key Laboratory for Research in Galaxies and Cosmology, Department of Astronomy, University of Science and Technology \\of China, Hefei, Anhui 230026, China\\
$^{6}$INAF, Osservatorio di Radioastronomia, Via Gobetti 101, I-40129, Bologna, Italy\\
$^{7}$Jodrell Bank Centre for Astrophysics, School of Physics and Astronomy, The University of Manchester, Oxford Road, \\Manchester M13 9PL, U.K. \\
$^{8}$Department of Physics \& Astronomy, Tufts University, 574 Boston Avenue, Medford, MA 02155, USA \\
$^{9}$Department of Physics, University of Zagreb, Bijeni\u{c}ka cesta 32, HR-10000 Zagreb, Croatia\\
$^{10}$Max Planck Institute for Astronomy, K\"{o}nigstuhl 17, Heidelberg D-69117, Germany}
\date{Accepted XXX. Received YYY; in original form ZZZ}

\pubyear{2016}

\begin{document}
\def\simlt{\mathrel{\rlap{\lower 3pt\hbox{$\sim$}}\raise 2.0pt\hbox{$<$}}}
\def\simgt{\mathrel{\rlap{\lower 3pt\hbox{$\sim$}} \raise
2.0pt\hbox{$>$}}}
\def\lsim{\,\lower2truept\hbox{${<\atop\hbox{\raise4truept\hbox{$\sim$}}}$}\,}
\def\gsim{\,\lower2truept\hbox{${> \atop\hbox{\raise4truept\hbox{$\sim$}}}$}\,}

\label{firstpage}
\pagerange{\pageref{firstpage}--\pageref{lastpage}}
\maketitle

\begin{abstract}
The assessment of the relationship between radio continuum luminosity and star formation rate (SFR) is of crucial importance to make reliable predictions for the forthcoming ultra-deep radio surveys and to allow a full exploitation of their results to measure the cosmic star formation history. We have addressed this issue by matching recent accurate determinations of the SFR function up to high redshifts with literature estimates of the 1.4\,GHz luminosity functions of star forming galaxies (SFGs). This was done considering two options, proposed in the literature, for the relationship between the synchrotron emission ($L_{\rm synch}$), that dominates at 1.4 GHz, and the SFR: a linear relation with a decline of the $L_{\rm synch}$/SFR ratio at low luminosities or a mildly non-linear relation at all luminosities. In both cases we get good agreement with the observed radio luminosity functions but, in the non-linear case, the deviation from linearity must be small. The luminosity function data are consistent with a moderate increase of the $L_{\rm synch}$/SFR ratio with increasing redshift, indicated by other data sets, although a constant ratio cannot be ruled out. A stronger indication of such increase is provided by recent deep 1.4\,GHz counts, down to $\mu$Jy levels. This is in contradiction with models predicting a \textit{decrease} of that ratio due to inverse Compton cooling of relativistic electrons at high redshifts. Synchrotron losses appear to dominate up to $z\simeq 5$.  We have also updated the \citet{Massardi2010} evolutionary model for radio loud AGNs.
\end{abstract}




\section{Introduction}\label{sec:intro}

In the last several years deep radio surveys reaching sub-mJy detection limits have been emerging as a powerful tool to investigate the evolution with cosmic time (or redshift) of star forming galaxies (SFGs), of Active Galactic Nuclei (AGNs) and of their mutual interactions. This is because the radio emission is not affected by dust obscuration and has straightforward K-corrections, thanks to the simple power-law shape of the spectra. On the other hand, the featureless radio spectra do not provide redshift information. This limitation has been overcome by  surveys down to tens of $\mu$Jy over fields with panchromatic coverage, such as the Cosmic Evolution Survey \citep[COSMOS;][]{Scoville2007} field and the Extended Chandra Deep Field-South \citep[E-CDFS;][]{Miller2013}.

The multi-frequency data available for these fields include spectroscopic redshifts for a substantial fraction of sources and have allowed accurate photometric redshift estimates for most of the others, as well as an effective separation of sources whose radio emission is due to star formation from those whose radio emission is powered by an active nucleus \citep{Smolcic2008, Ilbert2009, Sargent2010a, Bonzini2012, Bonzini2013}. With this information in hand, radio luminosity functions (RLFs) of both radio source populations have been derived up to high redshifts \citep{Smolcic2009a, Smolcic2009b, Padovani2011, Padovani2015, Novak17}.

Another investigation of the evolution of SFGs and of radio loud (RL) AGNs out to $z\sim 2.5$ was carried out by \citet{McAlpine2013} combining a 1 square degree Very Large Array (VLA) radio survey, complete to a depth of $100\,\mu$Jy, with accurate 10 band photometric redshifts from the Visible and Infrared Survey Telescope for Astronomy (VISTA) Deep Extragalactic Observations and
the Canada--France--Hawaii Telescope Legacy Survey. Less deep radio surveys are dominated by RL AGNs and were exploited to derive the RLFs of this population at several redshifts \citep{Donoso2009, Best2014}. Accurate estimates of the local RLFs of both SFGs and RL AGNs have been presented by \citet{MauchSadler2007}.

Identifications of sub-mJy radio sources have also revealed the presence of a third population, the `radio-quiet' (RQ) AGNs \citep[e.g.][]{Padovani2009, Bonzini2013}. Evidence of nuclear activity in these sources comes from one or more bands of the electromagnetic spectrum (e.g. optical, mid-infrared, X-ray) but the origin of their radio emission is still being hotly debated. High resolution radio observations have provided evidence that some RQ AGNs have components with very high surface brightness, variability, or apparent super-luminal motion, all of which are characteristic of emission driven by a super massive black hole \citep[e.g.,][]{HerreraRuiz2016, Maini2016}. 

However this does not necessarily prove that the general RQ AGN population possesses radio cores that contribute substantially to the total radio emission. Based on their study of the E-CDFS VLA sample \citet{Padovani2015} concluded that RQ and radio loud QSOs are two totally distinct AGN populations,  characterized by very different evolutions, luminosity functions, and Eddington ratios. The radio power of RQ AGNs evolves similarly to star-forming galaxies, consistent with their radio emission being powered by star formation. This conclusion was confirmed by the study of \citet{Bonzini2015} who used deep \textit{Herschel} photometry to determine the FIR emission, hence the SFR, of E-CDFS galaxies. Further support to this view was provided by \citet{Kellermann2016} who, based on 6 GHz Jansky Very Large Array (JVLA) observations of a volume-limited sample of 178 low redshift optically selected QSOs, argued that the bulk of the radio emission of RQ QSOs is powered by star formation in their host galaxies \citep[see also][]{Kimball2011, Condon2013}. In the following we will adopt this view and will not consider RQ AGNs as an additional radio source population.

This paper deals with the interpretation of these observational results and especially with their exploitation to assess the relationship between the 1.4\,GHz luminosity and the star formation rate (SFR) up to $z\simeq 5$. This relation is of crucial importance to make reliable predictions for the forthcoming ultra-deep surveys and to allow a full exploitation of their results to measure the cosmic star formation history. We have investigated it by comparing the observed RLFs at several redshifts with those expected from the accurate determinations of the star formation rate (SFR) functions of galaxies, converted to RLFs using literature relationships between SFR and radio (synchrotron and free-free) luminosity (Sect.~\ref{sec:SFmodel}). To have a comprehensive view of the outcome of radio surveys we need to take into account also RL AGNs. This is done using an updated version of the \citet{Massardi2010} model. In Sect.~\ref{sec:results} we discuss our results. The main conclusions are summarized in Sect.~\ref{sec:conclusions}. Throughout this paper we use a standard flat $\Lambda$CDM cosmology with $\Omega_{\rm m}=0.32$, $\Omega_\Lambda=0.68$, $h=0.67$.

\section{Modeling the evolution of the luminosity function of star-forming galaxies}\label{sec:SFmodel}

The radio continuum emission of SFGs is a well established SFR diagnostic, unaffected by dust obscuration \citep{KennicuttEvans2012}. It thus offers the opportunity to get a comprehensive view of the cosmic star formation history. In contrast, optical/UV surveys miss the heavily dust enshrouded star formation while the far-infrared (FIR) to millimeter-wave observations only measure the dust-reprocessed starlight. This is, in fact, one of the key science drivers of the Square Kilometer Array (SKA) and of its pathfinder telescopes, such as ASKAP and MeerKAT. To achieve this goal, however, it is necessary to set on a firm basis the connection between radio emission and SFR and to assess whether it depends on redshift.

The radio continuum emission of SFGs consists of a nearly flat-spectrum free-free emission plus a steeper-spectrum synchrotron component. The free-free emission is proportional to the production rate of ionizing photons, with a weak dependence on the electron temperature; thus it directly traces the formation rate of massive stars. The synchrotron emission, due to the interaction of relativistic electrons, mostly  produced by supernovae, with the galactic magnetic field, dominates at low radio frequencies.

A calibration of the relation between SFR and free-free emission was derived by \citet{Murphy2011} and \citet{Murphy2012},  based on observations of a sample of local galaxies, mostly with $\hbox{SFR}<1\,M_\odot\,\hbox{yr}^{-1}$. Following \citet{Mancuso2015b} we have rewritten it as:
\begin{equation} \label{eq:Lff}
 L_{\rm ff}(\nu)=3.75\times 10^{26} \left(\frac{\hbox{SFR}}{M_\odot/\hbox{yr}}\right) \, \left(\frac{T}{10^4\,\hbox{K}}\right)^{0.3}\!\!\!\!\!\! \hbox{g}(\nu,\hbox{T})\,\exp{\left(-{h\nu\over k\hbox{T}}\right)}
 \end{equation}
where $T$ is the temperature of the emitting plasma and  $\hbox{g}(\nu,\hbox{T})$ is the Gaunt factor for which we adopt the approximation given by \citet{Draine2011}
\begin{eqnarray}
\hbox{g}(\nu,\hbox{T})&=&\ln\left\{\exp\left[5.960-\frac{\sqrt{3}}{\pi}\ln\left(Z_i{\nu\over \hbox{GHz}} \left({T\over 10^4 \hbox{K}}\right)^{-1.5}\right)\right] \right.\nonumber \\
&+&\left.\exp(1)\right\},
\end{eqnarray}
$Z_i$ being the charge of ions. This equation reproduces the \citet{Murphy2012} calibration at the calibration frequency (33 GHz) for a pure hydrogen plasma ($Z_i=1$) and $T=10^4\,$K; we adopt these values in the following. The differences at other radio frequencies, due to the less accurate approximation for the Gaunt factor used by \citet{Murphy2012} are always $\simlt 3\%$.

The calibration of the SFR-synchrotron luminosity relation is a bit more controversial because it involves complex and poorly understood processes such as the production rate of relativistic electrons, the fraction of them that can escape from the galaxy, the magnetic field strength. A calibration of the $L_{\rm sync}$--SFR relation was obtained by \citet{Murphy2011,Murphy2012} using Starburst99 \citep{Leitherer1999} and observations of the low SFR, nearby galaxy samples mentioned above. Following \citet{Mancuso2015b} we have slightly modified their relation including a steepening of the synchrotron spectrum by $\Delta \alpha=0.5$ above a break frequency of 20 GHz, to take into account electron ageing effects \citep{BandayWolfendale1991}. The SFR-synchrotron luminosity relation then writes:
\begin{eqnarray}\label{eq:LsyncM}
L_{\rm sync}&\simeq &1.9\times 10^{28} \left(\frac{\hbox{SFR}}{\hbox{M}_{\odot}\hbox{yr}^{-1}}\right) \left(\frac{\nu}{\hbox{GHz}}\right)^{-0.85} \nonumber \\
&\cdot& \left[1+\left({\nu\over 20\rm GHz}\right)^{0.5}\right]^{-1}\, \hbox{erg}\,\hbox{s}^{-1}\,\hbox{Hz}^{-1}.
\end{eqnarray}
%
%
The calibration by \citet{Murphy2011,Murphy2012} was adopted in the extensively cited review by \citet{KennicuttEvans2012}. However, \citet{Mancuso2015b} showed that this relation, combined with observational determinations of the local SFR function, leads to an over-prediction of the faint end of the local RLF of SFGs worked out by \citet{MauchSadler2007}. A similar conclusion was previously reached by \citet{Massardi2010}.

This is in keeping with the argument made by \citet{Bell2003}. He pointed out that the FIR emission traces most of the SFR in luminous galaxies but only a minor fraction of it in faint galaxies. Nevertheless the FIR to radio luminosity ratio is similar for the two galaxy groups, implying that the radio emission from low-luminosity galaxies is substantially suppressed, compared to brighter galaxies.


\begin{figure}
  \hspace{+0.0cm} \makebox[0.48\textwidth][c]{
    \includegraphics[trim=2.7cm 0.5cm 1.0cm 0.5cm,clip=true,width=0.48\textwidth, angle=0]{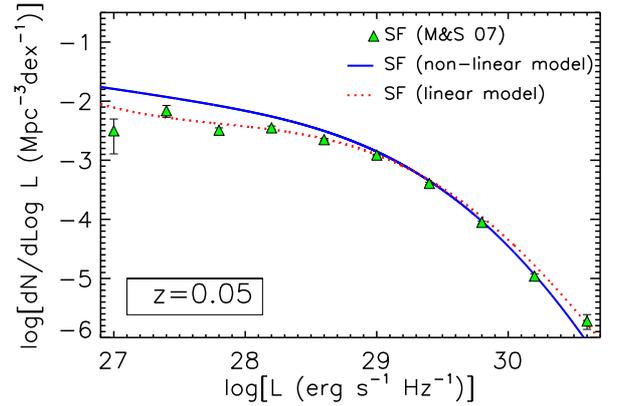}
  }
  \caption{Comparison between the observational determination of the local RLF of SFGs at 1.4 GHz \citep[data points;][]{MauchSadler2007} and the one obtained from the local SFR function using either eq.~(\ref{eq:Lsync_nonlin}) with a dispersion of 0.4 dex (dotted red curve) or eq.~(\protect\ref{eq:Lsync}) with a dispersion of 0.3 dex (solid blue line). }
  \label{fig:local_LF}
\end{figure}


\begin{figure}
  \hspace{+0.0cm} \makebox[0.48\textwidth][c]{
    \includegraphics[trim=2.7cm 0.5cm 1.0cm 0.5cm,clip=true,width=0.48\textwidth, angle=0]{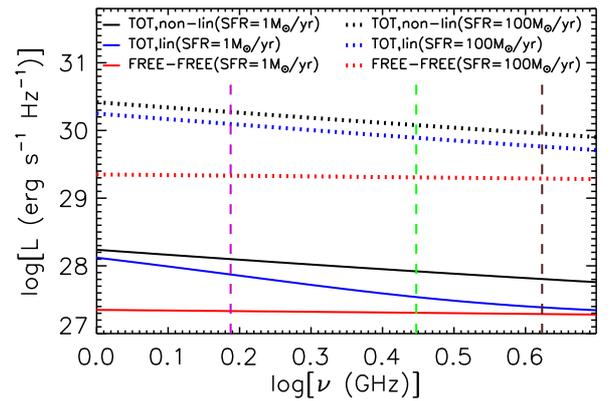}
  }
  \caption{Radio emission spectrum of SFGs for two values of the SFR (100 and  $1\,M_\odot\,\hbox{yr}^{-1}$, dotted and solid  lines, respectively). For each value of the SFR, the lowest dotted and solid red lines show the free-free contributions, while the two upper lines show the total, including the synchrotron contribution, computed using either eq.~(\ref{eq:Lsync}), uppermost black line of each type, or  eq.~(\ref{eq:Lsync_nonlin}), middle blue lines. The vertical dashed lines show the rest-frame frequencies corresponding to the observed frequency of 1.4\,GHz for $z=0.1$, 1, 2 (from left to right).  }
  \label{fig:spectrum}
\end{figure}

To recover consistency with the \citet{MauchSadler2007} LF \citet{Mancuso2015b} assumed a deviation from a linear $L_{\rm sync}$--SFR relation at low radio luminosities close to that proposed by \citet{Massardi2010}:
\begin{equation}\label{eq:Lsync_nonlin}
L_{\rm sync}(\nu)=\frac{L_{\star,\rm sync}(\nu)}{\left({L_{\star,\rm sync}}/{\bar{L}_{\rm sync}}\right)^{\beta}+\left({L_{\star,\rm sync}}/{\bar{L}_{\rm sync}}\right)},
\end{equation}
where $L_{\star,\rm sync} = 0.886\, \bar{L}_{\rm sync}({\rm SFR}=1\,M_\odot\,\hbox{yr}^{-1})$ with $\bar{L}_{\rm sync}$ given by eq.~(\ref{eq:LsyncM}), and $\beta = 3$; at 1.4\,GHz, $L_{\star,\rm sync}\simeq 10^{28}\, \hbox{erg}\,\hbox{s}^{-1}$ $\hbox{Hz}^{-1}$. A different empirical relation was proposed by \citet{Bell2003}.

The comparison with the \citet{MauchSadler2007} LF also showed that the bright end of the 1.4\,GHz LF is under-predicted unless we allow for a substantial dispersion around the mean ${L}_{\rm sync}$--SFR relation. \citet{Mancuso2015b} obtained good agreement adopting a dispersion of the synchrotron luminosity at given SFR $\sigma_{\log L}=0.4$ (see Fig.~\ref{fig:local_LF}). This dispersion is large compared to the much quoted value of 0.26 reported by \citet{Yun2001}, who however pointed out that the dispersion is larger at the high luminosities where it matters most in the present context. \citet[][their Table~5]{Murphy2012} found a scatter of 0.29 dex around the average ratio of the total IR luminosity (usually assumed to be a good SFR estimator) to the SFR inferred from 33 GHz measurements. Since the latter are dominated by thermal emission, that \citet{Murphy2012} argue to be a robust SFR indicator, it is not implausible that the dispersion around the mean ${L}_{\rm sync}$--SFR relation is somewhat larger. A larger dispersion (0.6 dex) of synchrotron luminosity, measured at 610\,MHz, and the SFR was reported by \citet{Garn2009} whose sample, however, included galaxies up to $z=2$; the dispersion may thus have been increased by evolution of the ${L}_{\rm sync}$--SFR relation. A dispersion of $\sim 0.4$ dex is also indicated by Figs.~6 and 9 of \citet{Delhaize17}.

Thus, although observational indications and theoretical arguments \citep[e.g.][]{Lacki2010} converge in suggesting deviations from a linear relation, its shape is uncertain. It is even unclear whether the ${L}_{\rm sync}$--SFR relationship becomes non-linear only at low radio luminosities. \citet{PriceDuric1992} and \citet{Niklas1997}, from analyses of samples of galaxies for which they were able to separate the free-free and the synchrotron components using the radio continuum spectra, concluded that their data are consistent with a mildly non-linear relation, holding at all radio luminosities: $L_{\rm sync}\propto \hbox{SFR}^{1.2}$.

We find that a relation of this kind yields a local RLF of SFGs in reasonably good agreement with the Mauch \& Sadler's one, although slightly steeper both above and below the knee luminosity. An excellent fit at high luminosities is obtained with a milder deviation from linearity ($L_{\rm sync}\propto \hbox{SFR}^{1.1}$); at low luminosities, however, the model is still above the data points. This is illustrated by Fig.~\ref{fig:local_LF} where the solid blue line, labeled ``non-linear model'', is obtained with a mean $L_{\rm sync}$/SFR relationship given by:
\begin{equation}\label{eq:Lsync}
\bar{L}_{\rm sync}\simeq 1.51\times 10^{28}\left(\frac{\nu}{\hbox{GHz}}\right)^{-0.85}\left(\frac{\hbox{SFR}}{\hbox{M}_{\odot}\hbox{yr}^{-1}}\right)^{1.1}\, \hbox{erg}\,\hbox{s}^{-1}\,\hbox{Hz}^{-1},
\end{equation}
and adopting a dispersion of $L_{\rm sync}$ at given SFR of $\sigma_{\log L}=0.3$. The normalization is consistent, within the uncertainties, with the \citet{PriceDuric1992} and \citet{Niklas1997} results. The dotted red line shows, for comparison, the fit with the ``linear model'', with deviations from linearity only at faint luminosities [eq.~(\ref{eq:Lsync_nonlin})]. The lower dispersion around the mean $L_{\rm sync}$--SFR relation obtained for the ``non-linear'', compared to the ``linear'' model suggests that the former might be preferred; on the other hand, its fit of the Mauch \& Sadler's local luminosity function is somewhat worse.

Figure~\ref{fig:spectrum} compares the radio emission spectra of SFGs yielded by eq.~(\ref{eq:Lsync_nonlin}) and by eq.~(\ref{eq:Lsync}) for 2 values of the SFR, 100 and  $1\,M_\odot\,\hbox{yr}^{-1}$. For both values of the SFR eq.~(\ref{eq:Lsync}) gives higher radio luminosities. Equation~(\ref{eq:Lsync_nonlin}) substantially lowers the synchrotron contribution for low SFRs to the point that the free-free emission of these sources may become dominant, even at 1.4 GHz (in the observer's frame), especially for high-$z$ sources. Hence determinations of the radio emission spectrum of SFGs with moderate to low SFRs ($\hbox{SFR}\simlt 1\,M_\odot\,\hbox{yr}^{-1}$), over a sufficiently broad frequency range would allow one to discriminate between the ``linear'' and the ``non-linear'' model.

All these studies however are limited to local galaxies with moderate luminosities/SFRs. Do their conclusions extend to high redshifts when SFRs reach much larger values? The applicability at high redshift of the local $L_{\rm sync}$--SFR relation was not firmly demonstrated yet. From a theoretical point of view it may be expected that the mean $L_{\rm sync}$/SFR ratio decreases at high redshifts when the cooling of relativistic electrons  via inverse Compton scattering off the cosmic microwave background, whose energy density increases as $(1+z)^4$, dominates over synchrotron cooling \citep[e.g.][]{Murphy2009, LackiThompson2010, SchleicherBeck2013, Schober2016}. On the other hand, under some circumstances the ratio might rather increase with redshift \citep{LackiThompson2010, SchleicherBeck2013}.

Observational studies concerned the relation between radio luminosity and FIR luminosity, generally believed to be a reliable measure of the SFR. In the relevant redshift range, the considered, relatively high, radio luminosities are dominated by synchrotron emission (cf. Fig.~\ref{fig:spectrum}), so that the conclusions concern, more specifically, the relation between synchrotron luminosity and SFR. The results have been controversial. Some studies found that the FIR-radio relation is unchanged or suffers only minor variations at high redshift \citep[e.g.][]{Ibar2008, Sargent2010b, Bourne2011, Mao2011, Pannella2015} while others have found significant, albeit weak, evolution \citep[e.g.][]{Seymour2009, Ivison2010a, Ivison2010b, Magnelli2015, Basu2015}. \citet{Magnelli2015}, based on FIR and radio observations of the most extensively studied extragalactic fields (GOODS-N, GOODS-S, ECDFS, and COSMOS) reported evidence of a weak redshift evolution of the parameter
\begin{equation}\label{eq:qfir}
q_{\rm FIR}=\log\left({L_{\rm FIR}[\hbox{W}]\over 3.75\times 10^{12}}\right)-\log\left(L_{1.4\,\rm GHz} [\hbox{W}\,\hbox{Hz}^{-1}]\right),
\end{equation}
{where $L_{\rm FIR}$ is the FIR luminosity integrated from rest-frame 42 to $122\,\mu$m and $L_{1.4\,\rm GHz}$ is the rest-frame 1.4\,GHz luminosity. They found}
\begin{equation}\label{eq:qevol}
q_{\rm FIR}=(2.35\pm 0.08)\times (1+z)^{\alpha_{\rm M}},
\end{equation}
{with $\alpha_{\rm M} = -0.12\pm 0.04$.}

{A new investigation of the evolution of the infrared-radio correlation has been recently carried out by \citet{Delhaize17}. Using highly sensitive 3 GHz  VLA observations and  \textit{Herschel}/\textit{Spitzer} infrared data in the $2\,\hbox{deg}^2$ COSMOS field they were able to push the study of the correlation up to $z\simeq 5$. Their results confirm a weak but statistically significant trend of $q_{\rm FIR}$ with redshift. The fitted redshift dependence of $q_{\rm FIR}$ has a slightly higher normalization and steeper slope than that found by \citet{Magnelli2015}: $q_{\rm FIR}=(2.88\pm 0.03)\times (1+z)^{-0.19\pm 0.01}$.}

These results are in striking contradiction with the inverse Compton scenario envisioned by several models, as mentioned above. A full investigation of this issue is beyond the scope of this paper. Various possibilities have been discussed by \citet{Magnelli2015} and \citet{Delhaize17}, including possible biases introduced by simplifying approximations used in their analyses. The choice of radio spectral index used for the K-correction was found to affect the normalization and the redshift dependence of $q_{\rm FIR}$: a steeper spectral index lowers the normalization and steepens the dependence on $(1+z)$. However, only marginal differences were found varying the spectral indices in the observed range. A flattening of the effective spectral index is produced by the increasing contribution of free-free emission toward higher radio frequencies; this translates in a flatter trend with $(1+z)$. However, a substantial flattening of the average effective radio spectral index with increasing redshift is inconsistent with their data.

The rapid increase with redshift of the AGN population hosted by SFGs might suggest that the evolution of $L_{\rm synch}$/SFR ratio is due to an increasing contribution of radio-quiet AGNs to the radio luminosity. To weed out the AGN contamination \citet{Magnelli2015} removed from their sample X-ray AGNs and AGNs selected by the \textit{Spitzer} Infrared Array Camera (IRAC) colour-colour criteria of \citet{Lacy2007}. Excluding AGNs did not change their results; they concluded that the evolution is most likely not driven by AGN contamination. The analysis by \citet{Delhaize17} led to a more open conclusion: it is possible that AGN contributions only to the radio regime could be influencing the trend found for their sample.

A physically-based prediction of the negative evolution of $q_{\rm FIR}$ was put forward by \citet{LackiThompson2010}. They argued that high-$z$ starburst galaxies have much higher cosmic ray vertical scale heights than local starbursts ($h\sim 1\,$kpc instead of $h\sim 0.1\,$kpc). Then energy losses of cosmic rays (free-free, ionization, pion production), competing with synchrotron cooling, are weaker because they depend on the volume density. If, moreover, the magnetic field strength, $B$, increases with the SFR surface density, $\Sigma_{\rm SFR}$, cosmic rays lose most of their energy via synchrotron emission before escaping from the galaxy and suffering inverse Compton cooling. Taking also into account that the other competing cooling processes are weaker it is then expected that  $q_{\rm FIR}$ decreases at high redshifts. According to \citet{Magnelli2015}, this scenario might account for the observed trend, although some difficulties remain.



We investigate the possible evolution of $q_{\rm FIR}$ (or, equivalently, of the $L_{\rm sync}$/SFR ratio) combining the accurate determinations of the redshift-dependent SFR functions that are presently available with the information on the RLFs of SFGs provided by the deep surveys mentioned in Sect.~\ref{sec:intro}. Detailed studies of the evolution of the SFR function across the cosmic history were carried out by \citet{Cai2013} focussing on FIR data and by \citet{Cai2014} focussing on UV and Ly$\alpha$ data, properly corrected for dust attenuation. The latter study, however, was limited to $z\gtrsim 2$. These authors built a model that fits a broad variety of data\footnote{See \url{http://people.sissa.it/~zcai/galaxy_agn/} or  \url{http://staff.ustc.edu.cn/~zcai/galaxy_agn/index.html}.}: multi-frequency and multi-epoch luminosity functions of galaxies and AGNs, redshift distributions, number counts (total and per redshift bins). It also accurately predicted the counts and the redshift distribution of strongly lensed galaxies detected by the South Pole Telescope \citep[SPT;][]{Mocanu2013,Weiss2013} as shown by \citet{Bonato2014}.

The model was extended by \citet{Mancuso2015b} and further successfully tested against observational determinations of the H$\alpha$ luminosity function at several redshifts. The combination of dust extinction corrected  UV/Ly$\alpha$/H$\alpha$ data with FIR data yielded accurate determinations of the SFR function up to $z\simeq 6$--7, some estimates being available up to $z\sim 10$ \citep[see also][]{Aversa2015}. 




\begin{figure*}
\hspace{+0.0cm} \makebox[\textwidth][c]{
  \includegraphics[trim=4.5cm -0.2cm 5.0cm 0cm,clip=true,width=0.9\textwidth, angle=0]{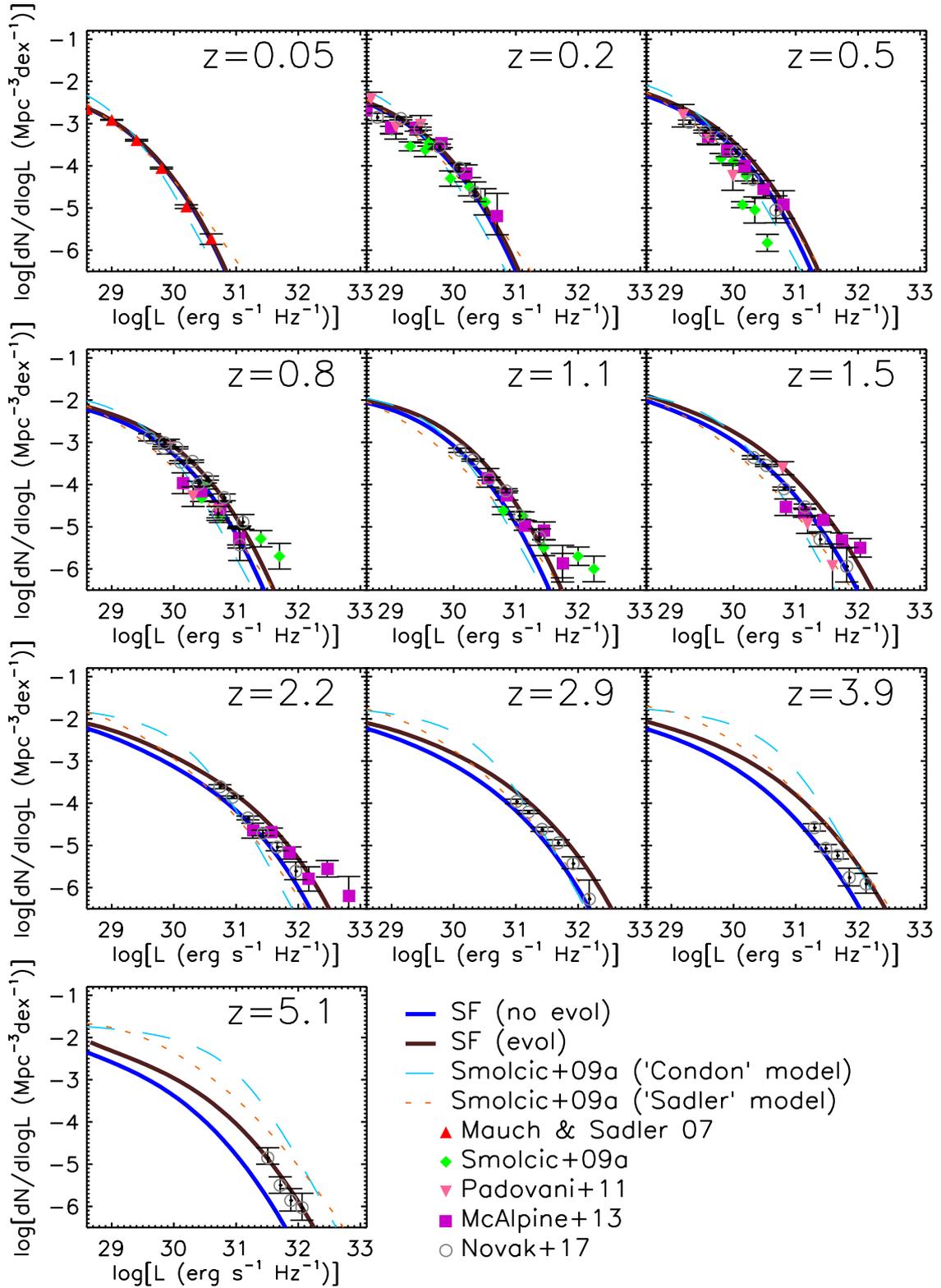}
  }
   \vspace{-0.5cm}
  \caption{Model RLFs (at 1.4 GHz) of SFGs compared with observational determinations specified in the inset. { The solid blue lines show the model RLFs inferred from the SFR functions, summing the free-free [eq.~(\ref{eq:Lff})] and the synchrotron contribution for the ``linear'' model [eq.~(\ref{eq:Lsync_nonlin})], without evolution of the mean $L_{\rm synch}$/SFR ratio. The solid dark-brown lines show the same model but with the best-fit evolution of the mean $L_{\rm synch}$/SFR ratio from \citet{Magnelli2015}.} Also shown, for comparison, are the \citet[][`Condon' model, dashed cyan lines]{Condon1989} and the \citet[][`Sadler' model, orange short dashes]{Sadler2002} parameterizations of the RLF \citep{Smolcic2009a}. }
  \label{fig:LF_catalogue_comp_SF}
\end{figure*}


\begin{figure*}
  \hspace{+0.0cm} \makebox[\textwidth][c]{
     \includegraphics[trim=4.5cm -0.3cm 5.0cm 0.0cm,clip=true,width=0.9\textwidth, angle=0]{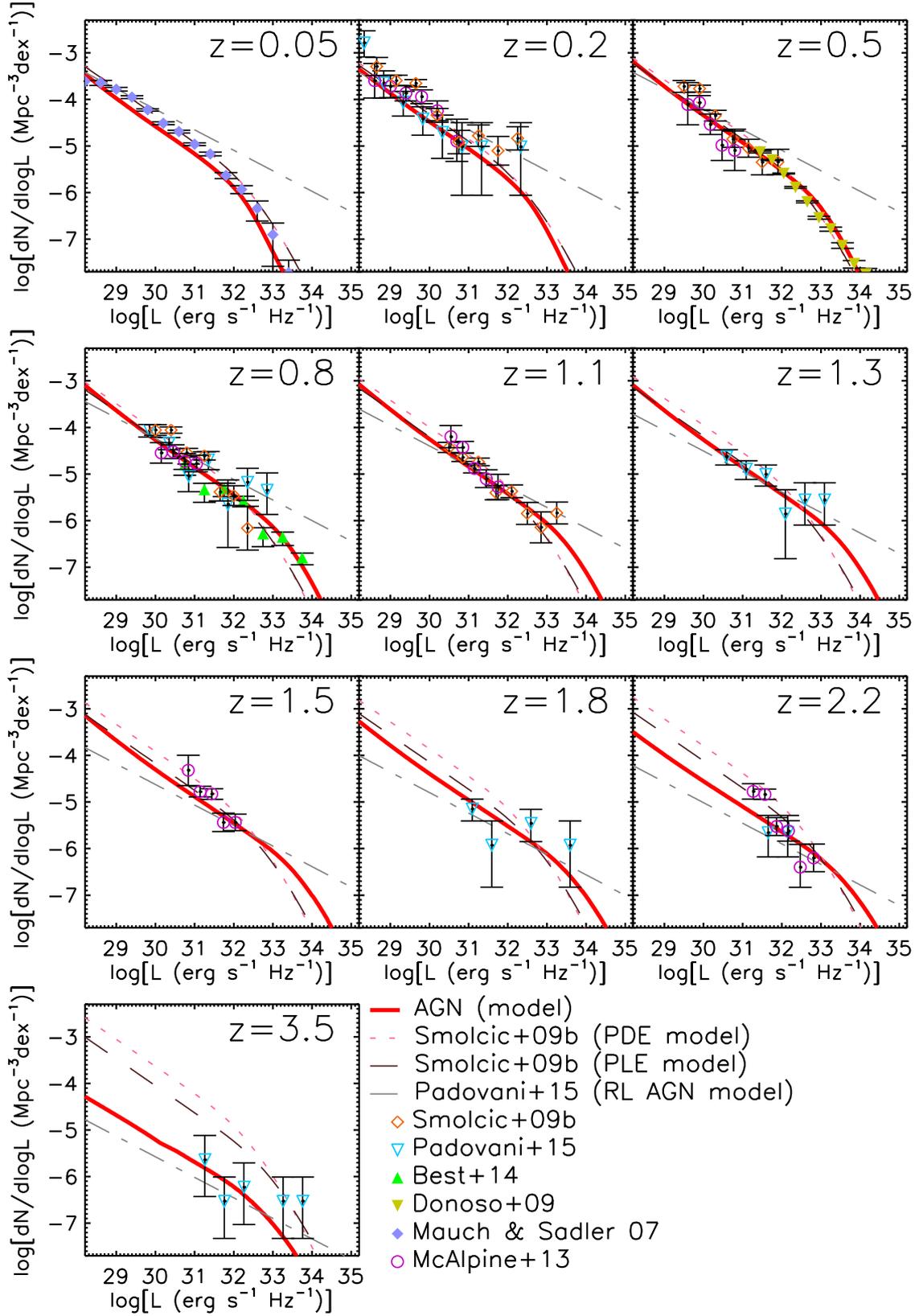}
  }
  \vspace{-0.5cm}
  \caption{RLFs (at 1.4 GHz) of RL AGNs yielded by the model described in Appendix~\ref{sec:RL} (solid red lines) compared with the observational estimates specified in the inset. Also shown, for comparison, are the \citet{Smolcic2009b} pure density evolution (PDE,  pink short dashes) and pure luminosity evolution (PLE, dashed brown lines) models, and the best-fitting `$z_{\rm peak}$ density evolution' (ZDE) model with single power-law LF of \citet[][dot-dashed grey lines]{Padovani2015}. }
  \label{fig:LF_catalogue_comp_AGN}
\end{figure*}


\begin{figure*}
  \hspace{0.0cm} \makebox[\textwidth][c]{
   \includegraphics[trim=4.7cm 0.7cm 5cm 1.0cm,clip=true,width=0.9\textwidth, angle=0]{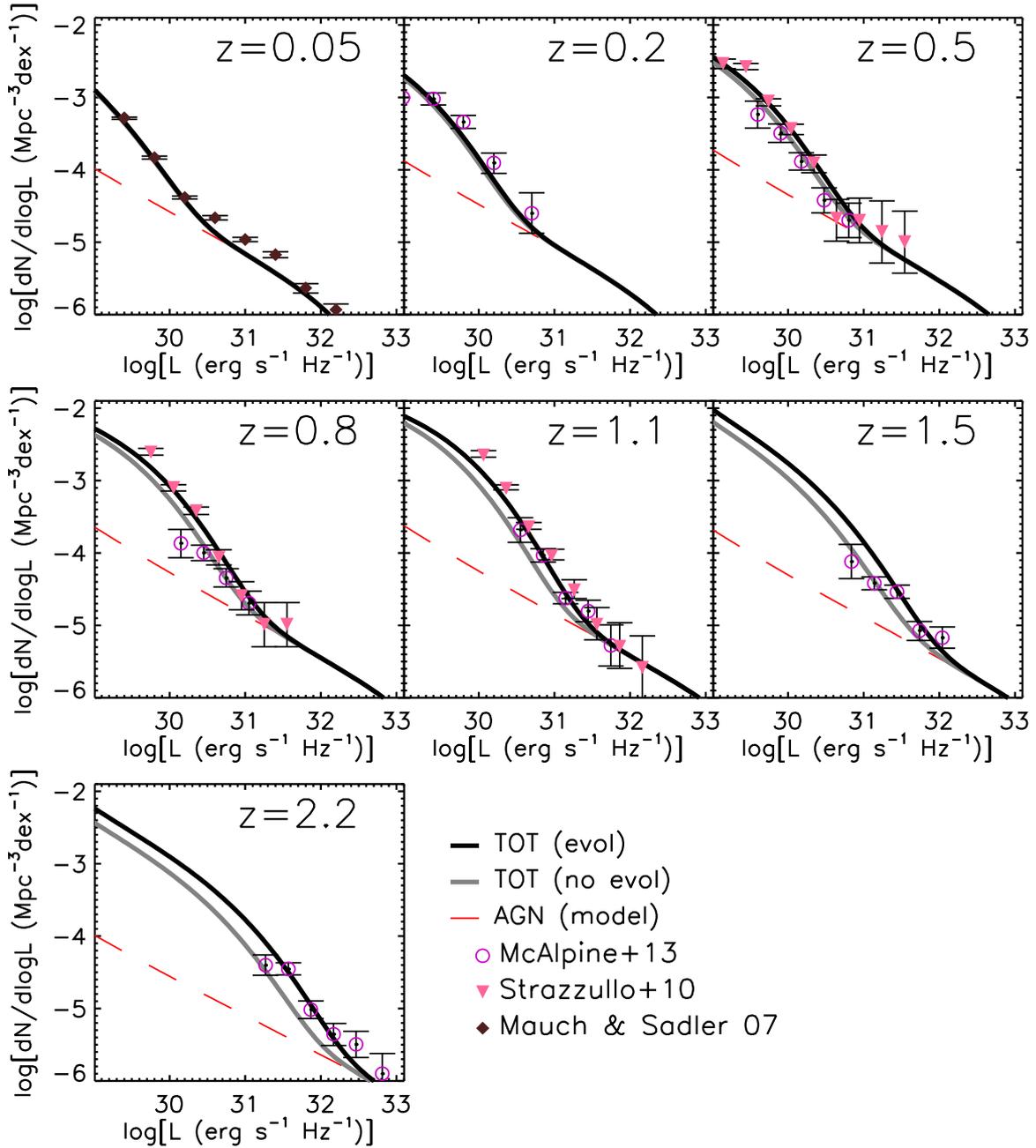}
  }
  \caption{{Total (SFG plus RL AGN) RLFs (at 1.4 GHz) derived using the ``linear'' model for the synchrotron emission, with and without evolution of the mean $L_{\rm synch}$/SFR ratio (solid black and grey lines, respectively), compared with estimates from the literature at several redshifts. The dashed red lines show the contributions of RL AGNs.} }
  \label{fig:LF_catalogue_comp_tot}
\end{figure*}


\begin{figure}
  \hspace{0.0cm} \makebox[0.48\textwidth][c]{
   \includegraphics[trim=2.5cm 0.55cm 1cm 0.3cm,clip=true,width=0.48\textwidth, angle=0]{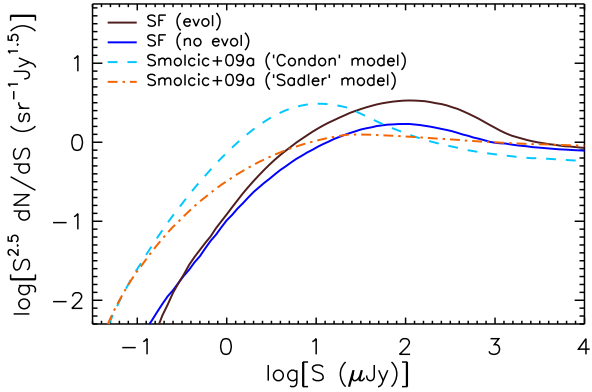}
  }
  \caption{{SFG counts yielded by the ``linear'' model for the synchrotron emission, with and without evolution of the mean $L_{\rm synch}$/SFR ratio (solid brown and blue line, respectively), compared with those given by the empirical models from \citet{Smolcic2009a}:  the `Condon' model (dashed cyan line) and the `Sadler' model (dot-dashed orange line).} }
  \label{fig:counts_comp}
\end{figure}


\begin{figure}
  \hspace{+0.0cm} \makebox[0.48\textwidth][c]{
    \includegraphics[trim=3.2cm 0.6cm 1.3cm 0.5cm,clip=true,width=0.48\textwidth, angle=0]{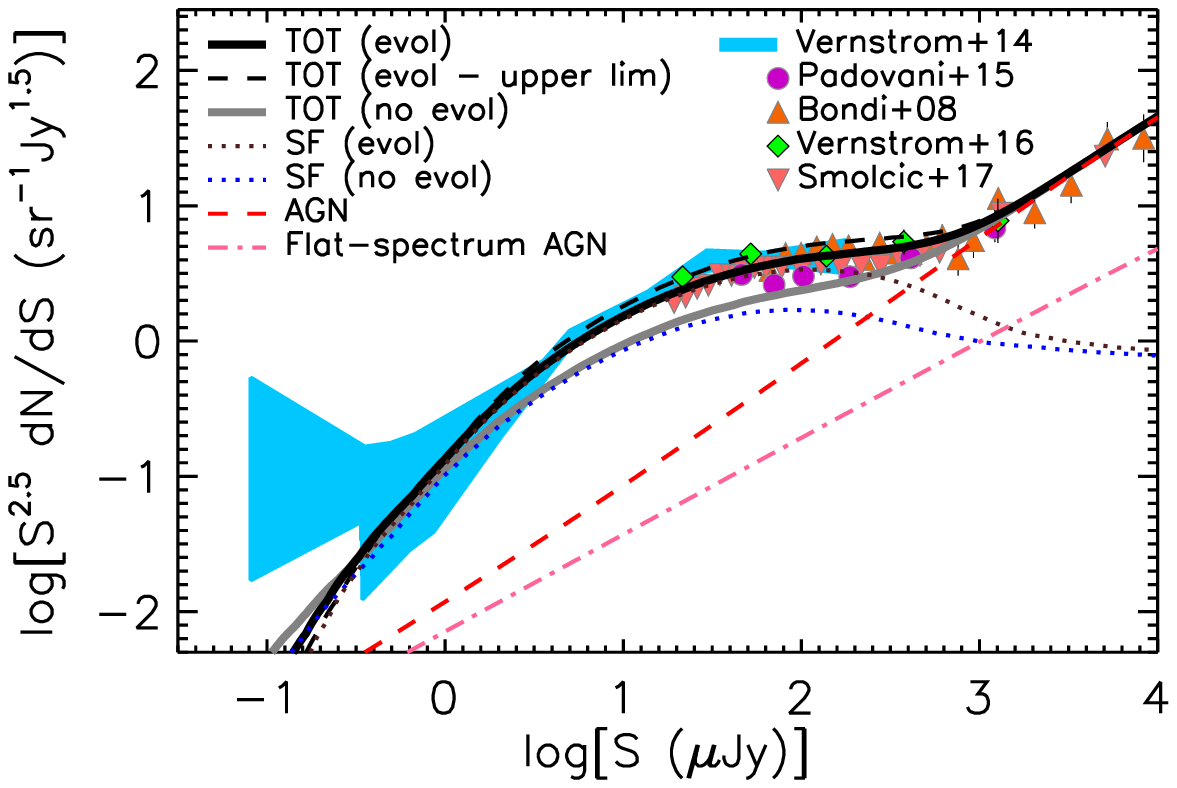}
  }
  \caption{Comparison between the observed faint 1.4\,GHz Euclidean normalized counts and those yielded by the model. The counts of RL AGNs (dashed red line) are dominated by steep-spectrum sources; the contribution of the flat-spectrum ones is shown by the dot-dashed pink line. The dotted lines show the counts of SFGs yielded by the ``linear'' model with (upper line) and without evolution of the mean $L_{\rm synch}$/SFR ratio, for the best fit value of the evolutionary parameter given by \citet{Magnelli2015}. The total counts for the two cases, obtained adding the RL AGN contribution, are shown by the solid black and grey lines, respectively. The evolution brings the model counts to agreement with the observational estimate by \citet{Bondi2008}, by \citet{Vernstrom2016} and by \citet{Smolcic2017}, and with the results of the $P(D)$ analysis by \citet[][shaded light blue area]{Vernstrom2014}. The ``non-linear'' model with slope of 1.1 yields counts (not shown) imperceptibly higher than those from the ``linear'' model. The dashed black line shows the model counts for an evolution of the mean $L_{\rm synch}$/SFR ratio at the $1\,\sigma$ limit of \citet{Magnelli2015}. In this case the counts are at the upper boundary of the observational determinations.
  }
  \label{fig:1d4GHz_counts}
\end{figure}

\section{Comparison between model and data}\label{sec:results}



Figure~\ref{fig:LF_catalogue_comp_SF} compares the model luminosity functions of SFGs obtained from the SFR functions using eq.~(\ref{eq:Lff}) plus eq.~(\ref{eq:Lsync_nonlin}) with the observational determinations at several redshifts available in the literature. Note that the model parameters were fixed to fit the \citet{MauchSadler2007} local luminosity function. No free parameters were introduced to fit the luminosity functions at higher redshifts.

The figure compares the luminosity functions obtained without evolution of the mean $L_{\rm synch}$/SFR ratio (solid blue lines) with those given by the same model but allowing for the best-fit evolution from \citet[][solid dark-brown lines]{Magnelli2015}. With these data alone it is hard to discriminate among the two possibilities, although the Magnelli et al. prediction provides a somewhat better description of the highest luminosity bins, especially at high redshift.


The main discrepancies between model (with and without evolution of the mean $L_{\rm synch}$/SFR ratio) and observational estimates, particularly with those by \citet{Smolcic2009a} at $z=0.8$ and $z=1.1$ and by \citet{McAlpine2013} at $z=2.2$, occur at the highest radio luminosities ($\log(L_{1.4\,\rm GHz}/\hbox{erg} \, \hbox{s}^{-1} \, \hbox{Hz}^{-1}) \simgt 31.5$) were the observational determinations tend to exceed expectations from the model. {To check whether this may be due, at least in part, to our model for the SFR function falling short at high luminosity we have replaced it with the analytic formulae given by \citet{Mancuso2016}. These formulae provide an accurate empirical fit of the observed SFR functions. The discrepancy was not significantly mitigated. } 

Note that the mean SFR of galaxies with 1.4\,GHz luminosity of $10^{31.5}\,\hbox{erg}\,\hbox{s}^{-1}\,\hbox{Hz}^{-1}$ is $\simeq 2800\,M_\odot\,\hbox{yr}^{-1}$ according to eq.~(\ref{eq:Lsync_nonlin}) or $\simeq 1355\,M_\odot\,\hbox{yr}^{-1}$ according to eq.~(\ref{eq:Lsync}). Even in the latter case, we are dealing with extreme SFRs that occur rarely. While the model predictions are, by construction, consistent with the space densities of galaxies with the corresponding SFRs, derived primarily from \textit{Herschel} data \citep{Gruppioni2013}, space densities derived from radio data are well above them. A contribution to the high luminosity excess may be due to strongly lensed SFGs. However, according to the estimates by \citet{Mancuso2015b}, this contribution is of order of 10\% or less in the relevant redshift and luminosity ranges.

Are radio surveys detecting SFGs with extreme SFRs, missed by FIR/sub-mm surveys? A more likely explanation is that the radio emission of galaxies with $\log(L_{1.4\,\rm GHz}/\hbox{erg} \, \hbox{s}^{-1} \, \hbox{Hz}^{-1}) \simgt 31.5$ has a substantial, perhaps dominant, contribution of nuclear origin. {A contamination of SFG samples by RL AGNs is not surprising. In fact, both the \citet{Smolcic2009a} and the \citet{McAlpine2013} SFG samples were built using an optical colour-based method to separate SFGs from AGNs. The latter were  identified in the radio population as sources associated to galaxies with predominantly old stellar populations, i.e. with redder colours. But studies of FIR properties of RL AGNs have found that their host galaxies show a strong evolution as a function of redshift, from predominantly quiescent at low $z$, except for the highest AGN powers \citep{Gurkan2015}, to predominantly dusty, with active star-formation at $z\simgt 1$ \citep[e. g.,][]{Kalfountzou2014, Magliocchetti2014, Magliocchetti2016, Rees2016}. Based on the results of \citet{McAlpine2013}, \citet{Magliocchetti2016} argue that the radio luminosity beyond which the radio emission is predominantly AGN-powered increases from  $\log(L_{1.4\,\rm GHz}/\hbox{erg} \, \hbox{s}^{-1} \, \hbox{Hz}^{-1}) \simeq 29.8$ at $z=0$ to $\log(L_{1.4\,\rm GHz}/\hbox{erg} \, \hbox{s}^{-1} \, \hbox{Hz}^{-1}) \simeq 31.6$ at $z=1.8$, and remains constant at higher redshifts. This is consistent with the luminosity ranges where the excess over the expected SFG contribution shows up (see Fig.~\ref{fig:LF_catalogue_comp_SF}). }

{A clear indication that the contamination of the bright tail of the RLFs of SFGs is due to  RL AGNs comes from Figs.~\ref{fig:LF_catalogue_comp_AGN} and \ref{fig:LF_catalogue_comp_tot}. The former figure shows that the observed space densities of the highest luminosity SFGs are akin to those of RL AGNs. RL AGNs are modeled following \citet[][see Appendix A for an upgraded version of such models]{Massardi2010}. The second shows that the total (SFG$+$RL AGN) RLFs above $\log(L_{1.4\,\rm GHz}/\hbox{erg} \, \hbox{s}^{-1} \, \hbox{Hz}^{-1}) \simeq 31.5$ can indeed be entirely accounted for by RL AGNs. 
}

{Strong support to {this conclusion} comes from the very recent analysis of the VLA-COSMOS 3\,GHz survey data by \citet{Novak17}. The increased sensitivity of the 3\,GHz survey compared to the 1.4\,GHz survey used by \citet{Smolcic2009a} and the different method used to select SFGs yielded $\sim 10$ times more SFG detections in the same redshift range. The new, statistically more robust determinations of the SFG RLFs do not show indications of excesses over the model predictions at any redshift (cf. Fig.~\ref{fig:LF_catalogue_comp_SF}).}

The ``non-linear'' model with slope 1.1 [eq.~(\ref{eq:Lsync})] yields luminosity functions hardly distinguishable from those of the ``linear'' model; therefore we chose not to plot them. The use of either eq.~(\ref{eq:Lsync_nonlin}) or eq.~(\ref{eq:Lsync}) yields reasonably good matches of the data. The quality of the fit looks quite similar by eye. A $\chi^2$ test gives a slight preference to the linear model, but the possibility of systematic errors due to, e.g., contamination by nuclear emission or errors on photometric redshift estimates, makes it difficult to reach a definite conclusion. However, the quality of the fit worsens rapidly as the slope on the $L_{\rm synch}$/SFR relation increases. Already for a slope of 1.2, the value suggested by \citet{PriceDuric1992}, the $\chi^2$ increases by more than a factor of 2. Thus, although a deviation from linearity at all luminosities cannot be ruled out, the data on the luminosity functions of SFGs imply that the slope cannot be significantly larger than 1.1.

Note that a contamination by RL AGNs, that flattens the brightest portion of the RLFs, mimics the effect of a slope larger than 1: the flattening could be misinterpreted as evidence for a ``non-linear'' relation. In the presence of a significant AGN contamination the minimum $\chi^2$ associated to the ``non-linear'' model would be underestimated and the upper limit to the slope would be overestimated.

As shown by Fig.~\ref{fig:LF_catalogue_comp_SF}, the available estimates of the RLFs of SFGs cover limited luminosity ranges. At $z> 1$ they are restricted to luminosities well above the RLF ``knee'' i.e. well above the luminosity, $L_\ast$, below which the slope of the RLF flattens substantially. But the largest contribution to the counts at any flux density $S$ comes from redshifts at which $L(S,z)\sim L_\ast$. It follows that the counts constrain the redshift-dependent SFG luminosity function at luminosities below those at which direct determinations have been obtained. The sensitivity of the counts to the detailed shape of the luminosity function is illustrated by Fig.~\ref{fig:counts_comp}: models that fit similarly well the available luminosity function data yield markedly different counts.

Figure~\ref{fig:1d4GHz_counts} compares  the observed 1.4\,GHz Euclidean normalized counts below 10\,mJy with those yielded by the model for RL AGNs and for SFGs. The counts of RL AGNs include the contributions of both steep- and flat-spectrum sources, the former being the dominant population. The counts of SFGs computed using the non-evolving ``linear'' model are lower than those by \citet{Bondi2008}, by \citet{Vernstrom2016} {and by \citet{Smolcic2017}}, and than the estimate based on the $P(D)$ analysis by \citet{Vernstrom2014}. The ``non-linear'' model with a slope of 1.1 (not shown) yields counts barely higher; the difference is almost indiscernible by eye.

Much better agreement with the observed counts is obtained  with the ``linear'' model, adopting the evolution of the $L_{\rm synch}$/SFR ratio based on the results of \citet[][cfr. eqs.~(\ref{eq:qfir}) and (\ref{eq:qevol})]{Magnelli2015} (solid black line in Fig.~\ref{fig:1d4GHz_counts}):  
\begin{equation}\label{eq:Magnelli}
\log L_{\rm synch,1.4\,\rm GHz}(z)=\log L_{\rm synch,1.4\,\rm GHz}(0) +2.35[1-(1+z)^{-0.12}].
\end{equation}
$L_{\rm synch,1.4\,\rm GHz}(0)$ is given by eq.~(\ref{eq:Lsync_nonlin}) and $L_{\rm FIR}$ has been converted to SFR using the calibration by \citet{KennicuttEvans2012}.  The source counts thus lend support to the case for an increase with $z$ of the $L_{\rm synch}$/SFR ratio, for which the \citet{Magnelli2015} data yielded a significance at the $3\,\sigma$ level only. A comparison with Fig.~\ref{fig:counts_comp} shows that such good fit, obtained without playing with any free parameter, is a highly non-trivial result.

{The dashed black line in Fig.~\ref{fig:1d4GHz_counts} illustrates the sensitivity of the counts to the evolution of the $L_{\rm synch}$/SFR ratio. Already with the coefficient of eq.~(\ref{eq:qfir}) and $\alpha_{\rm M}$ at their $1\,\sigma$ limits (2.43 and $-0.16$, respectively) the evolution yields counts at hundreds of $\mu$Jy flux densities at the upper limits of  observational determinations. The ``non-linear'' model with slope 1.1 [eq.~(\ref{eq:Lsync})] gives counts very close to those of the latter case, i.e. requires a slightly weaker evolution to match the counts. Thus the counts, while supporting the evolution of the $L_{\rm synch}$/SFR ratio, also provide tight constraints on its strength.}

\section{Conclusions}\label{sec:conclusions}

We have exploited the observational estimates of the 1.4\,GHz RLFs of SFGs at several redshifts, up to $z\simeq 5$, to investigate the relationship between the 1.4\,GHz luminosity and the SFR over a wider luminosity and redshift range than done so far. The assessment of such a relationship is crucial on one hand to enable a full exploitation of surveys from the SKA and its precursors to measure the cosmic star formation history and on the other hand to provide reliable predictions for future deep radio surveys. To this end we have exploited recent accurate determinations of the SFR functions up to high redshifts obtained by combining data from optical, UV, H$\alpha$ and FIR/sub-mm surveys.

At the low radio frequencies ($\simlt 5\,$GHz) at which deep large area surveys are preferentially carried out or planned, the dominant emission process is synchrotron radiation, whose connection to the SFR is not well understood, although very well established observationally. Several observational indications and theoretical arguments point to deviations from a linear relation between the synchrotron emission, $L_{\rm synch}$, and SFR. However the parameters defining such a relation are fraught with a considerable uncertainty. It is even unclear whether the deviations from linearity are limited to low luminosities/SFRs \citep{Bell2003, Lacki2010, Massardi2010, Mancuso2015b} or extend to all luminosities \citep{PriceDuric1992,Niklas1997,Basu2015}.

We have derived 1.4\,GHz luminosity functions from the SFR functions both in the case of deviations from a linear relation limited to low luminosities (``linear'' model) and in the case of a mildly non-linear relation at all luminosities (``non-linear'' model). In both cases the model parameters were fixed fitting the local luminosity function of SFGs by \citet{MauchSadler2007}.

The quality of the fit of observational determinations of the 1.4\,GHz luminosity function at several redshifts is similarly good in both cases. A $\chi^2$ analysis slightly favours the ``linear'' model, but we must be aware of systematics that are difficult to quantify. Thus with the data at hand we cannot clearly discriminate among the two options. However the RLF data reject ``non-linear'' models with a slope $\ge 1.2$.

The ``linear'' and ``non-linear'' options differ most markedly at the highest and at the lowest luminosities. Current surveys are not deep enough to reach sufficiently low luminosities at high redshifts to allow for a clear discrimination. At the highest luminosities the observed luminosity functions of SFGs may be contaminated by nuclear radio emission and must therefore be dealt with caution. Much larger and deeper samples, and better ancillary data to allow a more solid classification are necessary to accurately determine the parameters of the $L_{\rm sync}$/SFR relation.

We have modeled the source counts with two components, one accounting for SFGs and one for RL AGNs (i.e. we have assumed that the radio emission from RQ AGNs is dominated by star formation). Under this assumption, both the ``linear'' and the ``non-linear'' cases, without evolution of the $L_{\rm synch}$/SFR ratio, yield sub-mJy counts  substantially below the observational determinations by \citet{Bondi2008}, \citet{Vernstrom2016} and \citet{Smolcic2017}, and the $P(D)$ estimate from \citet{Vernstrom2014}. Good consistency is achieved, for the ``linear'' case, adopting the best fit evolution derived by \citet{Magnelli2015}. A similarly good fit is obtained, for the ``non-linear'' case, with a slightly weaker evolution. This strengthens the case for evolution that \citet{Magnelli2015} found to be significant only at the $3\,\sigma$ level. At the same time, the counts provide a tight upper limit to the strength of the evolution. Already the $1\,\sigma$ limits of the \citet{Magnelli2015} evolutionary parameters yield counts at the upper boundary of observational determinations at hundreds of $\mu$Jy flux densities. 

The increase with redshift of the $L_{\rm synch}$/SFR ratio, supported by our results, is in contradiction with models predicting a \textit{decrease} of such ratio  at high redshifts as a consequence of the increase of the energy losses of relativistic electrons via inverse Compton scattering off the cosmic microwave background \citep[e.g.][]{Murphy2009, LackiThompson2010, SchleicherBeck2013, Schober2016}.


As for RL AGNs, we find good consistency between the observational determinations of the redshift-dependent RLFs and our update of the \citet{Massardi2010} model, without any adjustment of the parameters. This lends support to the description of downsizing built in the model.


\section*{Acknowledgements}
We are grateful to Marco Bondi for useful clarifications on the VLA--COSMOS catalogue. The comments of the anonymous referees have helped us to better focus our paper and to clarify several points. Work supported in part by PRIN--INAF 2014 ``Probing the AGN/galaxy co-evolution through ultra-deep and ultra-high resolution radio surveys'', by  PRIN--INAF 2012 ``Looking into the dust-obscured phase of galaxy formation through cosmic zoom lenses in the Herschel Astrophysical Large Area Survey'' and by ASI/INAF agreement n. 2014-024-R.1. Z.Y.C. is supported by the National Science Foundation of China (grant No. 11503024). A.B. acknowledges support from the European Research Council under the EC FP7 grant number 280127. V.S. acknowledges the European Union's Seventh Framework programme under grant agreement 337595. P.C. thanks the support of the Ministry of Foreign Affairs and International Cooperation, Directorate General for the Country Promotion (Bilateral Grant Agreement ZA14GR02 - Mapping the Universe on the Pathway to SKA). M.N. acknowledges financial support from the European Union's Horizon 2020 research and innovation programme under the Marie Sk{\l}odowska-Curie grant agreement No 707601.

\bibliographystyle{mnras}
\bibliography{radioLF} 

\appendix

\section{Evolution of radio loud AGNs}\label{sec:RL}

To describe the cosmological evolution of RL AGNs we adopted the \citet{Massardi2010} model, which successfully fitted a large amount of data on LFs of steep- and flat-spectrum sources, multi-frequency source counts and redshift distributions. The model includes two flat-spectrum populations with different evolutionary properties (flat-spectrum radio quasars, FSRQs, and BL Lacs) and a single steep-spectrum population. For sources of each population a simple power-law spectrum is adopted: $S\propto \nu^{-\alpha}$, with $\alpha_{\rm FSRQ}=\alpha_{\rm BLLac}=0.1$, and $\alpha_{\rm steep} = 0.8$.

The epoch-dependent comoving LFs (in units of ${\rm Mpc^{-3}\, (d\log L)^{-1}}$) were modeled as double power-laws:
\begin{equation}\label{eq:Phi}
\Phi(L(z), z)=\frac{n_0}{(L(0)/L_\star(0))^a+(L(0)/L_\star(0))^b}{d\log L(0)\over d\log L(z)}.
\end{equation}
\citet{Massardi2010} worked out, for each population, an analytic luminosity evolution model entailing a high-$z$ decline of the comoving LF:
\begin{equation}\label{eq:evol}
L_{\star}(z)\!=\! L_{\star}(0)\,{\rm dex}\!{\left[k_{\rm evo}z\!\left(\!2z_{\rm top}\!-\!{2z^{m_{\rm ev}}z_{\rm top}^{(1-m_{\rm ev})}}\!/(1\!+\!m_{\rm ev})\right)\right]},
\end{equation}
where $z_{\rm top}$ is the redshift at which $L_{\star}(z)/ L_{\star}(0)$ reaches its maximum {and $|m_{\rm ev}|\le 1$}. The data further required a luminosity dependence of $z_{\rm top}$, that was parameterized as
\begin{equation}\label{eq:ztop}
z_{\rm top}=z_{{\rm top},0}  + {\delta z_{\rm top}\over 1+ L_{\star}(0)/L}.
\end{equation}
The fit gave positive values for the parameters $z_{{\rm top},0} $ and $\delta z_{\rm top}$, implying that the high-$z$ decline of the space density is more pronounced and starts at lower redshifts for less powerful sources. This behaviour is qualitatively similar to the {\it downsizing} observed for galaxies and optically and X-ray selected quasars. We included the Jacobian term
\begin{eqnarray}\label{eq:Jacobian}
\frac{d\log L(0)}{d\log L(z)}&=&1+2k_{\rm evo}\delta z_{\rm top}\ln10\frac{L_{*}(0)/L(0)}{(1+[L_{*}(0)/L(0)])^{2}}\nonumber \\ &-&\frac{2\ln10(1-m_{\rm ev})}{1+m_{\rm ev}}\cdot\delta z_{\rm top}\cdot\frac{L_{*}(0)}{L(0)}\cdot \nonumber \\
&\cdot&\!\!\!\!\frac{\left[z_{\rm top,0}+\delta z_{\rm top}/(1+L_{*}(0)/L(0))\right]^{-m_{\rm ev}}}{(1+L_{*}(0)/L(0))^{2}},
\end{eqnarray}
inadvertently omitted in the code used by \citet{Massardi2010}. We have recomputed the best fit values of the parameters using the same method and comparing the model to the same data sets. The global minimum $\chi^2$ decreases by about 10\% and the best fit values of the parameters change a little. The new values are listed in Table~\ref{tab:parameters}.

\begin{table}
\centering
\footnotesize
\begin{tabular}{lccc}
\hline
\hline
Parameters & FSRQ & BLLac & SS-AGNs \\
\hline
$a$ &  0.743   &  0.786  &  0.487 \\
$b$ &  3.293   &  1.750  &  2.410 \\
$\log n_{0}$ &  -11.262   &  -7.683  & -5.866  \\
$\log L_{*}(0)$ & 34.285    &  33.223  & 32.472  \\
$k_{\rm evo}$ &  -0.976   & 0.582   & 1.244  \\
$z_{\rm top,0}$ &  1.749   &  1.054  &  1.063 \\
$\delta z_{\rm top}$ &  0.001   & $-$   &  0.772 \\
$m_{\rm ev}$ & -0.207    &  1  & 0.278  \\
\hline
\hline
\end{tabular}
\caption{Best-fit values of the parameters of the evolutionary model for RL AGNs (see eqs.~\ref{eq:Phi}, ~\ref{eq:evol}, ~\ref{eq:ztop} and ~\ref{eq:Jacobian}).}
\label{tab:parameters}
\end{table}

As found by \citet{Massardi2010}, the luminosity dependence of the peak redshift required by the data is substantial for the steep-spectrum population but is weak ($\delta z_{\rm top}\ll 1$) for FSRQs. This is expected since the contributions to the observables (source counts and redshift distributions) of moderate to low luminosity flat-spectrum sources are dominated by BL Lacs, so that the evolution of low luminosity FSRQs is poorly constrained. In the case of BL Lacs the data are not enough to constrain the parameters governing the luminosity dependence of the evolution. Thus, following \citet{Massardi2010}, for this population, we have set $m_{\rm ev}=1$ and $\delta z_{\rm top}= 0$. 

\begin{figure}
  \hspace{0.0cm} 
 \includegraphics[trim=0.0cm -0.4cm 1.0cm 0.0cm,clip=true,width=0.48\textwidth, angle=0]{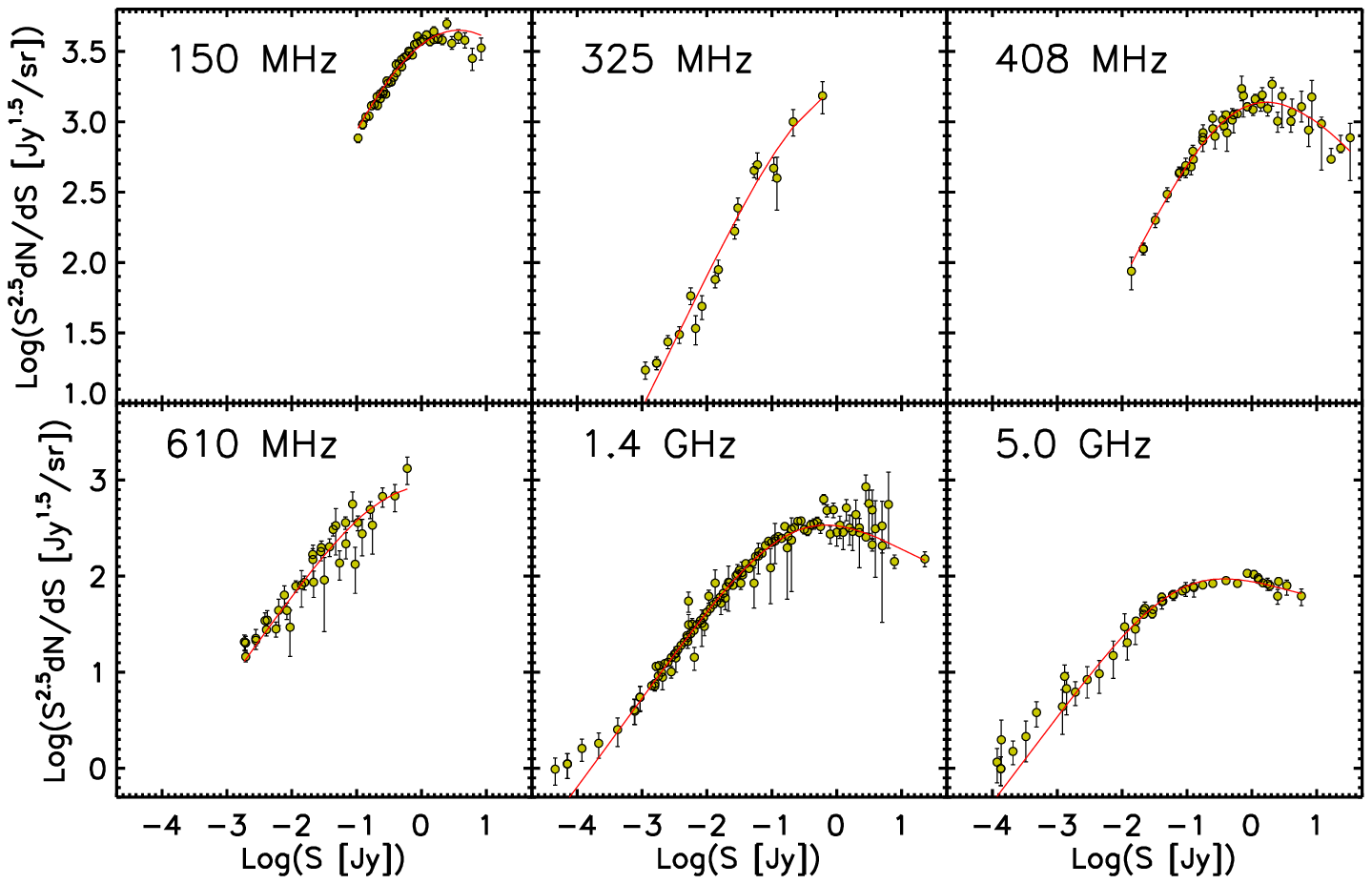}
  \caption{Comparison of our best-fit evolutionary model for RL AGNs with Euclidean normalized differential counts at several frequencies. Data points are from the compilation by \citet{DeZotti2010}. As is well known, AGNs become sub-dominant compared to SFGs at sub-mJy flux density levels. }
  \label{fig:RLcounts}
\end{figure}


\begin{figure}
  \hspace{0.0cm} 
 \includegraphics[trim=2.0cm -0.8cm 2.5cm 0.0cm,clip=true,width=0.48\textwidth, angle=0]{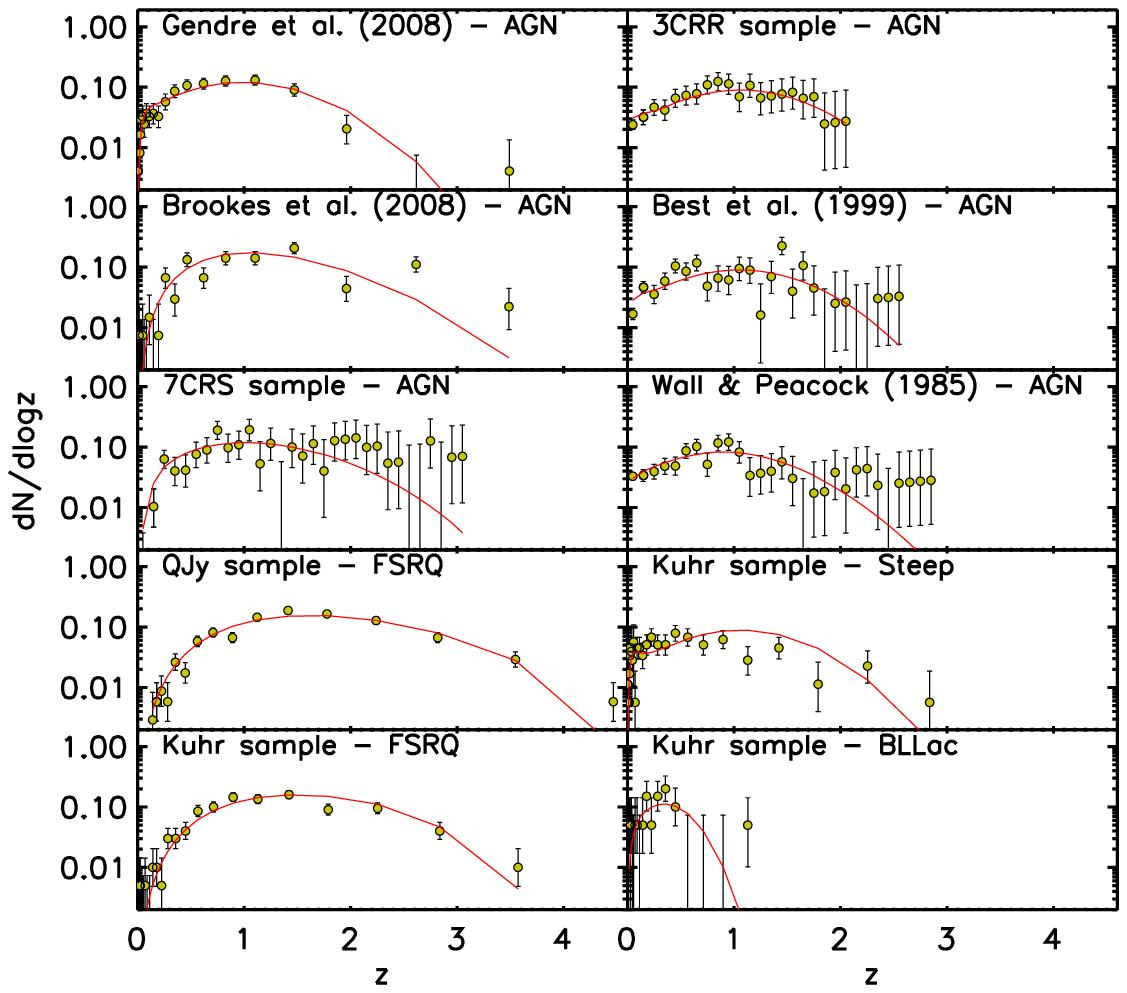}
  \caption{Comparison of our best-fit evolutionary model for RL AGNs with observational determinations of the redshift distributions for several source samples. See \citet{Massardi2010} for a description of the samples and for references.}
  \label{fig:RLzdistr}
\end{figure}

Figures~\ref{fig:RLcounts} and \ref{fig:RLzdistr} compare the outcome of the revised model with the observed source counts at different frequencies and with the observational determinations of the redshift distributions for 10 source samples at different frequencies, at different flux density limits and for different populations of RL AGNs. The agreement of the model with the data is always satisfactory.

Our model for RL AGNs is quite successful at accounting for the evolution of the RLFs of these objects (see Fig.~\ref{fig:LF_catalogue_comp_AGN}). As mentioned above, the dominant AGN population is made of steep-spectrum sources for which evidence of downsizing was inferred from integrated data (primarily source counts). The adopted description of the downsizing turns out to be nicely consistent with the more direct information provided by the redshift dependent luminosity functions.

Figures~\ref{fig:LF_catalogue_comp_SF} and \ref{fig:LF_catalogue_comp_AGN} also show that our model is highly competitive with the most recent models found in the literature. This is a non-trivial result because the other models shown were tuned to fit specifically the observational determinations of the 1.4\,GHz luminosity functions while our model is not. In fact, our redshift-dependent RLFs of SFGs are directly extrapolated from the SFR functions and those of RL AGNs come from fits of other data sets, without any adjustment of the parameters.

\bsp	
\label{lastpage}
\end{document}